\documentclass[3p,times,twocolumn]{elsarticle}
 \biboptions{comma,sort&compress}
 
\usepackage{graphicx}
\usepackage{here}
\usepackage{ecrc}

\volume{00}

\firstpage{1}

\journalname{Nuclear and Particle Physics Proceedings}

\runauth{}

\jid{nppp}

\jnltitlelogo{Nuclear and Particle Physics Proceedings}

\usepackage{amssymb}
\usepackage[figuresright]{rotating}

\newcommand{\be}{\begin{equation}}

\newcommand{\ee}{\end{equation}}
\newcommand{\bea}{\begin{eqnarray}}
\newcommand{\eea}{\end{eqnarray}}
\newcommand{\beas}{\begin{eqnarray*}}
\newcommand{\eeas}{\end{eqnarray*}}

\newcommand{\nn}{\nonumber\\}

\newcommand{\me}{\mathrm{e}} %math e
\newcommand{\mi}{\mathrm{i}} %math i
\newcommand{\df}[1]{\ensuremath{\frac{\mathrm{d}^{3}#1}{(2\pi)^{3}}}\,}

\begin{document}

\begin{frontmatter}

%%
%%%%%%%%%%%%%%%%%%%%%%%%%%%%%%%%%%%%%%%%%%%%%%%%%
\title{The (magnetized) effective QCD phase diagram$^*$}
 % \corref{cor0}}
 \cortext[cor0]{Talk given at 18th International Conference in Quantum Chromodynamics (QCD 15,  30th anniversary),  29 june - 3 july 2015, Montpellier - FR}
 \author[label1,label2]{Alejandro Ayala}
%  \cortext[cor0]{FAPESP CNPq-Brasil PhD student fellow.}
\ead{ayala@nucleares.unam.mx}
\address[label1]{Instituto de Ciencias Nucleares, Universidad Nacional Aut\'onoma de M\'exico, 
Apartado Postal 70-543, M\'exico Distrito Federal 04510, M\'exico.}
\address[label2]{Centre for Theoretical and Mathematical Physics, and Department 
of Physics, University of Cape Town, Rondebosch 7700, South Africa.}
%\address[label2]{Laboratoire
%Particules et Univers de Montpellier, CNRS-IN2P3, 
%Case 070, Place Eug\`ene
%Bataillon, 34095 - Montpellier, France.}
% \author[label3]{F. Fanomezana\corref{cor1}}
%  \cortext[cor1]{PhD student.}
%\ead{fanfenos@yahoo.fr}
%\address[label3]{Institute of High-Energy Physics of Madagascar (iHEP-MAD), University of Antananarivo, 
%Madagascar}
% \author[label2,label4]{S. Narison\fnref{fn1}}
%   \fntext[fn1]{Speaker, Corresponding author.}
%    \ead{snarison@yahoo.fr}
%
%\address[label4]{Madagascar consultant of the Abdus Salam International Centre for Theoretical Physics (ICTP), via Beirut 6,34014 Trieste, Italy .}
% \author[label3]{A. Rabemananjara\corref{cor1}}
%%  \cortext[cor2]{Ph.D. student}
%\ead{achris\_01@yahoo.fr}

\pagestyle{myheadings}
\markright{ }
\begin{abstract}
I present the highlights of a recent study of the effective QCD phase diagram on the temperature $T$ and
quark chemical potential $\mu$ plane, where the strong interactions are modeled 
using the linear sigma model coupled to quarks. The phase
transition line is found from the effective potential at finite
$T$ and $\mu$ taking into account the plasma screening
effects. We find the location of the critical end point (CEP) to
be $(\mu^{\mbox{\tiny{CEP}}}/T_c,T^{\mbox{\tiny{CEP}}}/T_c)\sim(1.2,0.8)$,
where $T_c$ is the (pseudo)critical temperature for the crossover
phase transition at vanishing $\mu$. This location lies within the
region found by lattice inspired calculations. Since the linear sigma model does not exhibit confinement, I argue that the location is due to the proper treatment of the plasma screening effects and not to the size of the confining scale. I also comment on the extension of this study
to determine the dependence of the CEP's location on the strength of an external magnetic field.
\end{abstract}
% \begin{document}

\end{frontmatter}
%%%%%%%%%%%%
%\vspace*{-1.5cm}
\section{Introduction}

The different phases in which matter made up of quarks and
gluons arranges itself depends on
the temperature and density, or equivalently, on the temperature
and chemical potentials. The representation of the QCD
phase diagram is thus two dimensional. This is customary plotted with the
light-quark chemical potential $\mu$ as the horizontal variable
and the temperature $T$ as the vertical one.  $\mu$ is related to
the baryon chemical potential $\mu_B$ by $\mu_B=3\mu$.

Most of our knowledge of the phase diagram is restricted to the
$\mu=0$ axis. The phase diagram is, by and large, unknown. For
physical quark masses and $\mu=0$, lattice calculations have
shown~\cite{Aoki} that the change from the low temperature phase,
where the degrees of freedom are hadrons, to the high temperature
phase described by quarks and gluons, is an analytic crossover.
The phase transition has a dual nature: on the one hand the
color-singlet hadrons break up leading to deconfined quarks and
gluons; this is dubbed as the {\em deconfinement phase
transition}. On the other hand, the dynamically generated
component of quark masses within hadrons vanishes; this is
referred to as {\em chiral symmetry restoration}.

Lattice calculations have provided values for the crossover
(pseudo)critical temperature $T_c$ for $\mu=0$ and 2+1 quark
flavors using different types of improved rooted staggered
fermions. The MILC collaboration
obtained $T_c=169(12)(4)$ MeV. The RBC-Bielefeld
collaboration reported $T_c=192(7)(4)$
MeV. The Wuppertal-Budapest
collaboration has consistently obtained
smaller values, the latest being $T_c=147(2)(3)$ MeV. The HotQCD
collaboration has computed $T_c=154(9)$ MeV and 
more recently $T_c=155(1)(8)$ MeV~\cite{lattice}. The
differences could perhaps be attributed to different lattice spacings.

Although the above picture presented by lattice QCD  cannot
be easily extended to the case $\mu\neq 0$ due to the {\it sign problem}, some mathematical extensions of
lattice techniques~\cite{mathlattice} as well as Schwinger-Dyson equations~\cite{Roberts}, can  be employed
to explore all the phase diagram.

A number of different model approaches indicate
that the transition along the $\mu$ axis, at $T=0$, is strongly
first order~\cite{first-order}. Since the first order line
originating at $T=0$ cannot end at the $\mu=0$ axis which
corresponds to the starting point of the cross-over line, it must
terminate somewhere in the middle of the phase diagram. This point
is generally referred to as the critical end point (CEP). The mathematical extensions of
lattice techniques place the CEP in the region
$(\mu^{\mbox{\tiny{CEP}}}/T_c,T^{\mbox{\tiny{CEP}}}/T_c)\sim(1.0-1.4,0.9-0.95)$~\cite{Sharma}.

In the first of Refs.~\cite{Roberts}, it is argued that the
theoretical location of the CEP depends on the size of the
confining length scale used to describe strongly interacting
matter at finite density/temperature. This argument is supported
by the observation that the models which do not account for this
scale~\cite{NJL,pNJL,Chqm,pChqm} produce either a CEP closer to
the $\mu$ axis ($\mu^{\mbox{\tiny{CEP}}}/T_c$ and
$T^{\mbox{\tiny{CEP}}}/T_c$ larger and smaller, respectively) or a
lower $T_c$~\cite{nNJL} than the lattice based approaches or the
ones which consider a finite confining length scale. Given the
dual nature of the QCD phase transition, it is interesting to
explore whether there are other features in models which have
access only to the chiral symmetry restoration facet of QCD that,
when properly accounted for, produce the CEP's location more in
line with lattice inspired results.

An important clue is provided by the behavior of the critical
temperature as a function of an applied magnetic field. Lattice
calculations have found that this temperature decreases when the
field strength increases~\cite{Fodor,Bali:2012zg,Bali2}. It has
been recently shown that this phenomenon, dubbed {\it inverse
magnetic catalysis}, can be obtained in models, such as  the Abelian Higgs
model or the linear sigma model with quarks, which show only chiral symmetry restoration and lack confinement.
This result is a consequence of the decrease of the coupling constants
with increasing field strength. The novel feature
implemented in these calculations is the handling of the screening
properties of the plasma, which effectively makes the treatment go
beyond the mean field approximation~\cite{Ayala1, Ayala2} and allows to consider the thermomagnetic
modifications of the coupling constants at lowest order within the same calculation. Screening is also important to
obtain a decrease of the coupling constant with the magnetic field strength in QCD in the Hard Thermal Loop
approximation~\cite{Ayala3}. 

It therefore seems that properly accounting for the plasma screening effects in effective models allows to obtain
both a CEP's location in line with lattice inspired techniques as well as inverse magnetic catalysis. A pertinent question is what happens to the CEP's location once a magnetic field dependence is included in the analysis for $\mu\neq 0$ and wether the nature of the phase transition for $\mu=0$ changes as the magnetic field strength increases. Recent lattice QCD calculations~\cite{Endrodimag} show that at very high values of the magnetic field strength, inverse magnetic catalysis prevails and that the phase transition becomes first order at asymptotically large values of the magnetic field for $\mu=0$ (see also Ref.~\cite{Costa}). In this work we explore the
consequences of the proper handling of the plasma screening
properties in the description of the effective QCD phase diagram
within the linear sigma model with quarks. We argue that it is the adequate description of these
properties which determines the CEP's location. 
We find that for certain values of the model parameters, obtained from physical
constraints, the CEP's location agrees with lattice inspired calculations. We also give a preview of work in progress~\cite{Ayala4} that shows that when including the effects of a magnetic field in the calculation of both the effective potential as well as on the thermomagnetic dependence of the coupling constants, the CEP's location moves toward smaller values of the chemical potential and lower temperatures and that above a certain value of the field strength the CEP reaches the $T$-axis and the phase transitions become first order, also in line with recent lattice results~\cite{Endrodimag}. Details of the calculation that forms the basis of this work can be found in Ref.~\cite{Ayala5}.

\section{The linear sigma model with quarks}

We start from the linear sigma model coupled to quarks. It is
given by the Lagrangian density
\begin{eqnarray}
   {\mathcal{L}}&=&\frac{1}{2}(\partial_\mu \sigma)^2  + \frac{1}{2}(\partial_\mu \vec{\pi})^2 + \frac{a^2}{2} (\sigma^2 + \vec{\pi}^2) \nonumber \\
   &-& \frac{\lambda}{4} (\sigma^2 + \vec{\pi}^2)^2 
   + i \bar{\psi} \gamma^\mu \partial_\mu \psi \nonumber \\
   &-& g\bar{\psi} (\sigma + i \gamma_5 \vec{\tau} \cdot \vec{\pi} )\psi,
\label{lagrangian}
\end{eqnarray}
where $\psi$ is an SU(2) isospin doublet, $\vec{\pi}=(\pi_1,
\pi_2, \pi_3 )$ is an isospin triplet and $\sigma$ is an isospin
singlet. The neutral pion is taken as the third component of the
pion isovector, $\pi^0=\pi_3$ and the charged pions as
$\pi_\pm=(\pi_1\mp i\pi_2)/2$. The squared mass parameter $a^2$
and the self-coupling $\lambda$ and $g$ are taken to be positive.

To allow for the spontaneous breaking of symmetry, we let the
$\sigma$ field develop a vacuum expectation value $v$ 
\bea
   \sigma \rightarrow \sigma + v,
\label{shift} 
\eea 
which can later be taken as the order parameter
of the theory.  After this shift, the Lagrangian density can be
rewritten as 
\bea
   {\mathcal{L}} &=& -\frac{1}{2}\sigma\partial_{\mu}\partial^\mu\sigma-\frac{1}
   {2}\left(3\lambda v^{2}-a^{2} \right)\sigma^{2}\nn
   &-&\frac{1}{2}\vec{\pi}\partial_{\mu}\partial^\mu\vec{\pi}-\frac{1}{2}\left(\lambda v^{2}- a^2 \right)\vec{\pi}^{2}+\frac{a^{2}}{2}v^{2}\nn
  &-&\frac{\lambda}{4}v^{4} + i \bar{\psi} \gamma^\mu \partial_\mu \psi
  -gv \bar{\psi}\psi + {\mathcal{L}}_{I}^b + {\mathcal{L}}_{I}^f,
  \label{lagranreal}
\eea
where ${\mathcal{L}}_{I}^b$ and  ${\mathcal{L}}_{I}^f$ are given by
\begin{eqnarray}
  {\mathcal{L}}_{I}^b&=&-\frac{\lambda}{4}\Big[(\sigma^2 + (\pi^0)^2)^2\nn
  &+& 4\pi^+\pi^-(\sigma^2 + (\pi^0)^2 + \pi^+\pi^-)\Big],\nn
  {\mathcal{L}}_{I}^f&=&-g\bar{\psi} (\sigma + i \gamma_5 \vec{\tau} \cdot \vec{\pi} )\psi,
  \label{lagranint}
\end{eqnarray}
and describe the interactions among the fields $\sigma$,
$\vec{\pi}$ and $\psi$, after symmetry breaking. From
Eq.~(\ref{lagranreal}) we see that the $\sigma$, the three pions
and the quarks have masses 
\bea
  m^{2}_{\sigma}&=&3  \lambda v^{2}-a^{2},\nn
  m^{2}_{\pi}&=&\lambda v^{2}-a^{2}, \nn
  m_{f}&=& gv,
\label{masses}
\eea
respectively.

The one-loop effective potential for the linear sigma model with quarks 
including the plasma screening properties encoded in the ring diagrams contribution has been calculated in detail for zero chemical potential in Refs.~\cite{Ayala:2009ji, Ayala:2014mla}. Such analyses show that inclusion of the ring 
diagrams renders the effective potential stable. 
%It is given by
%\bea
%V^{({\mbox{\small{eff}}})}&=&
%-\frac{a^2}{2}v^2 + \frac{\lambda}{4}v^4\nn
%&+&\sum_{i=\sigma,\vec{\pi}}\left\{\frac{m_i^4}{64\pi^2}
%\left[ \ln\left(\frac{(4\pi T)^2}{2a^2}\right)\right.\right. \nn
%&-&\left. 2\gamma_E +1\frac{}{}\right] 
%- \frac{\pi^2T^4}{90} + \frac{m_i^2T^2}{24}\nn
%&-&\left. \frac{T}{12 \pi}(m_i^2 + \Pi)^{3/2} \right\} \nn
%&-& N_{c}\sum_{f=u,d}\left\{\frac{m_f^4}{16\pi^2}
%\left[\ln\left(\frac{(\pi T)^2}{2a^2}\right)\right.\right.\nn
%&-&\left. 2\gamma_{E}+1\right] \left. -\frac{m_f^2T^2}{12} + \frac{7\pi^{2}T^{4}}{180} \right\},
%\label{Veff-mu0}
%\eea
%where $\Pi=\frac{\lambda T^2}{2}+\frac{N_fN_cg^2T^2}{6}$ is the 
%self-energy for any of the bosons. 

When the $\mu$ is non-vanishing, the calculation of the effective
potential is more complicated. Though the boson contribution 
remains the same, the fermion contribution has to be modified due to the chemical potential. 
The modification enters the calculation in two ways: 
indirectly into the boson self-energy and 
directly from its contribution to the effective potential.

To one-loop order the fermion contribution to the effective potential in the
imaginary time formalism of thermal field theory is given 
by~\cite{Ayala:2009ji}
\begin{eqnarray}
V_{f}&\!\!\!\!=\!\!\!\!&-\frac{2}{\beta}\int\df{k}
\left[\beta\omega + \ln\left(1+\me^{-\beta(\omega-\mu)}\right)\right. \nn
&\!\!\!\!+\!\!\!\!& \left. \ln\left(1+\me^{-\beta(\omega+\mu)} \right) \right],
\label{Vf}
\end{eqnarray}
where $\beta=T^{-1}$ and $\omega=(\vec{k}^{2}+m_{f}^{2})^{1/2}$, and
the sum over the fermion Matsubara frequencies has been performed. The first term in Eq.~(\ref{Vf}) corresponds to the vacuum contribution whereas the second and third ones are the matter contributions. Note that the matter contribution is made out of separate quark and antiquark pieces due to the finite
chemical potential. The vacuum contribution is well-known~\cite{Ayala:2009ji} and can be expressed, after mass renormalization as a function of the renormalization scale $\tilde{\mu}$. For the evaluation of the medium's contribution in Eq.~(\ref{Vf}) we adapt the technique from Ref.~\cite{Jackiw} to the 
present case. The main idea is to produce a
second-order differential equation in $y^{2}$, where $y=m_{f}/T$, valid at high temperature with
$m_{f}$ as the smallest of all scales, for the finite temperature part of the 
potential, which we denote by $\tilde{V}_{f}$, given in 
Eq.~(\ref{Vf}) with appropriate boundary conditions at $y=0$, where the integrals can be analytically evaluated. The expression for the effective potential is obtained by 
integrating this differential equation and using the given boundary conditions. 
%The second-order differential equation satisfied by $\tilde{V}_{f}$ is
%\begin{eqnarray}
%\frac{d^{2}\tilde{V}_{f}}{dy^{4}}&\!\!\!\!=\!\!\!\!&\frac{1}{8\pi^{2}\beta^{4}}
%\left[\ln\left(\frac{y^{2}}{(4\pi)^{2}}\right)
%-\psi^{0}\left(\frac{1}{2}+\frac{\mi z}{2\pi}\right) \right. \nn
%&\!\!\!\!-\!\!\!\!& \left.\psi^{0}\left(\frac{1}{2}-\frac{\mi z}{2\pi}\right)\right],
%\label{diffeqn}
%\end{eqnarray}
%where $\psi^0(x)$ is the digamma function. The boundary conditions 
%are
%\begin{eqnarray}
%%\label{bc1}
%\tilde{V}_{f}|_{y^{2}=0}&\!\!\!\!=\!\!\!\!&
%\frac{2}{\pi^{2}\beta^{4}}\left[Li_4(-e^{z})+Li_4(-e^{-z})\right]\nn
%%\label{bc2}
%\frac{d\tilde{V}_{f}}{dy^{2}}|_{y^{2}=0}&\!\!\!\!=\!\!\!\!&
%\frac{-1}{2\pi^{2}\beta^{4}}\left[Li_2(-e^{z})+Li_2(-e^{-z})\right],
%\label{boundarycond}
%\end{eqnarray}
%where $Li_n(x)$ is the polylogarithm function of order $n$. 
%
%The boundary conditions Eqs.~(\ref{boundarycond}) fix the two 
%integration constants needed to determine $\tilde{V}_{f}(y,z)$.
%The solution of Eq.~(\ref{diffeqn}) that satisfies the boundary conditions
%Eqs.~(\ref{boundarycond}) is given by
%\begin{eqnarray}
%\tilde{V}_{f}&\!\!\!\!=\!\!\!\!&-\frac{1}{16\pi^{2}\beta^{4}}
%\left\{-y^{4}\ln\left(\frac{y^{4}}{(4\pi)^{2}}\right)
%\right. \nn
%&\!\!\!\!+\!\!\!\!& y^{4}\left[\frac{3}{2} + \psi^{0}\left(\frac{1}{2}+\frac{\mi z}{2\pi}\right)
%+\psi^{0}\left(\frac{1}{2}-\frac{\mi z}{2\pi}\right)\right] \nn
%&\!\!\!\!+\!\!\!\!& 8y^{2}\left[Li_2(-e^{z})+Li_2(-e^{-z})\right] \nn
%&\!\!\!\!-\!\!\!\!& \left. 32\left[Li_4(-e^{z})+Li_4(-e^{-z})\right]\right\}.
%\label{Vfexplpre}
%\end{eqnarray}
Combining the vacuum contribution after mass renormalization with the 
finite temperature part 
%and recalling that $y=m_{f}/T$ and $z=\mu/T$
we finally have~\cite{Ayala5}
\begin{eqnarray}
\tilde{V}_{f}&\!\!\!\!=\!\!\!\!&-\frac{1}{16\pi^{2}}
\left\{ m_{f}^{4}\left[\ln\left(\frac{(4\pi T)^{2}}{2\tilde{\mu}^{2}}\right) 
\right.\right. \nn
&\!\!\!\!+\!\!\!\!& \left. \psi^{0}\left(\frac{1}{2}+\frac{\mi\mu }{2\pi T}\right)
+ \psi^{0}\left(\frac{1}{2}-\frac{\mi\mu}{2\pi T}\right)\right] \nn
&\!\!\!\!+\!\!\!\!& 8m^{2}T^{2}\left[Li_2(-e^{\mu/T})+Li_2(-e^{-\mu/T})\right] \nn
&\!\!\!\!-\!\!\!\!& \left. 32T^{4}\left[Li_4(-e^{\mu/T})+Li_4(-e^{-\mu/T})\right]\right\}.
\label{Vfexpl}
\end{eqnarray}

It can also be shown that the boson self-energy $\Pi$ 
computed for a finite chemical potential and in the limit where the 
masses are small compared to $T$, is given by
\bea
\Pi&=&\frac{\lambda T^2}{2}\nn
     &-&\frac{N_fN_cg^2T^2}{\pi^2}\left[ Li_2(-e^{\mu/T}) + Li_2(-e^{-\mu/T})\right],
\label{self-energy} 
\eea 
where $N_f=2$ and $N_c=3$ are the number of light flavors and colors,
respectively.

%**************************************************************************

Choosing the renormalization scale as $\tilde{\mu}=e^{-1/2}a$, the effective 
potential up to the ring diagrams contribution is then given by
\bea
   \!\!\!\!\!\!\!\!V^{({\mbox{\small{eff}}})}&\!\!\!\!=\!\!\!\!&
   -\frac{a^2}{2}v^2 + \frac{\lambda}{4}v^4\nn
  \!\!\!\! \!\!\!\!&\!\!\!\!+\!\!\!\!&\sum_{i=\sigma,\vec{\pi}}\left\{\frac{m_i^4}{64\pi^2}
   \left[ \ln\left(\frac{(4\pi T)^2}{2a^2}\right)
   \right.\right.\nn
   \!\!\!\!\!\!\!\!&\!\!\!\!-\!\!\!\!&\left. 2\gamma_E +1\frac{}{}\right]-\frac{\pi^2T^4}{90} + \frac{m_i^2T^2}{24}\nn
   \!\!\!\!\!\!\!\!&\!\!\!\!-\!\!\!\!& \left.\frac{T}{12 \pi}(m_i^2 + \Pi)^{3/2} \right\} \nn
   \!\!\!\!\!\!\!\!&\!\!\!\!-\!\!\!\!& \frac{N_{c}}{16\pi^2}\sum_{f=u,d}
   \left\{m_f^4\left[\ln\left(\frac{(4\pi T)^2}{2a^2}\right)+1\right.\right.\nn
   \!\!\!\!\!\!\!\!&\!\!\!\!+\!\!\!\!&\left.\psi^0\left(\frac{1}{2}+\frac{i\mu}{2\pi T}\right) +
          \psi^0\left(\frac{1}{2}-\frac{i\mu}{2\pi T}\right) \right] \nn
   \!\!\!\!\!\!\!\!&\!\!\!\!+\!\!\!\!&8\ m_f^2T^2\left[ Li_2(-e^{\mu/T}) + Li_2(-e^{-\mu/T})\right]\nn
   \!\!\!\!\!\!\!\!&\!\!\!\!-\!\!\!\!& \left. 32\ T^4 \left[ Li_4(-e^{\mu/T}) + Li_4(-e^{-\mu/T}) \right]\right\}.
\label{Veff-mid}
\eea

In the limit when $\mu\rightarrow 0$, Eq.~(\ref{Veff-mid}) becomes the expression found in Refs.~\cite{Ayala:2009ji, Ayala:2014mla}. In the same limit, Eq.~(\ref{self-energy}) reduces to the well known expression for the self-energy at
high temperature~\cite{Ayala2}. 
%Note that the self-energy provides the screening to render the
%effective potential in Eq.~(\ref{Veff-mid}) stable. Should this
%self-energy be absent, the term $(m_i^2 + \Pi)^{3/2}$ would
%instead be $(m_i^2)^{3/2}$, which becomes imaginary when for
%certain values of $v$,  $m_i^2$ becomes negative [see
%Eqs.~(\ref{masses})]. This term is obtained from considering the
%resummation of the ring diagrams and therefore
Equation~(\ref{Veff-mid}) represents the effective potential computed
beyond the mean field approximation that accounts for the leading
screening effects at high temperature.

\section{The phase diagram}

\begin{figure}[t!]
\begin{center}
\includegraphics[scale=0.85]{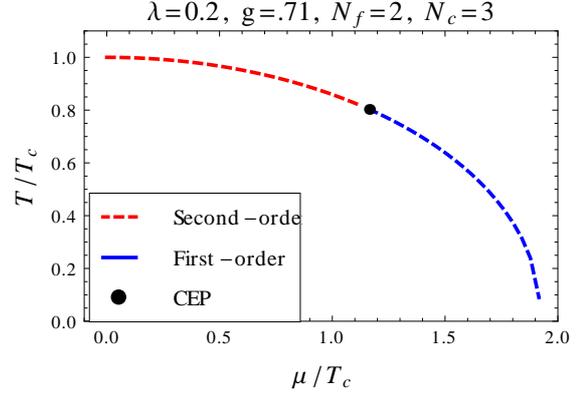}
\end{center}
\caption{Effective QCD phase diagram computed for $\lambda = 0.2$ and $g = 0.71$ obtained by considering $m_\sigma^{\mbox{\tiny{vac}}} = 300$ MeV. For small values of $\mu$
the phase transition is second order. The order of the transition
changes to first order for larger values of $\mu$. The CEP is
located at
$(\mu^{\mbox{\tiny{CEP}}}/T_c,T^{\mbox{\tiny{CEP}}}/T_c)\sim(1.2,0.8)$.}
\label{fig1}
\end{figure}

In order to find the values of the parameters $\lambda$, $g$ and $a$ appropriate for the description of the phase transition, we note that when considering the thermal effects the boson masses are modified since they acquire a thermal component. For $\mu=0$ they become
\bea
   m_\sigma^2(T)&=&3\lambda v^2 -a^2 + \frac{\lambda T^2}{2}+\frac{N_fN_cg^2T^2}{6}\nn
   m_\pi^2(T)&=&\lambda v^2 -a^2 + \frac{\lambda T^2}{2}+\frac{N_fN_cg^2T^2}{6}.
\label{massmod} 
\eea 
At the phase transition, the curvature of the
effective potential vanishes for $v=0$. Since the boson thermal
masses are proportional to this curvature, these also vanish at
$v=0$. From any of the Eqs.~(\ref{massmod}), we obtain a relation
between the model parameters at $T_c$ given by
\bea
   a=T_c\sqrt{\frac{\lambda}{2}+\frac{N_fN_cg^2}{6}}.
\label{relation}
\eea
Furthermore, we can fix the value of $a$ by noting from Eqs.~(\ref{masses}) that the vacuum boson masses satisfy
\bea
   a=\sqrt{\frac{m_\sigma^2 - 3m_\pi^2}{2}}.
\label{massvac}
\eea
Since in our scheme we consider two-flavor massless quarks in the chiral limit, we take $T_c\simeq 170$ MeV~\cite{example} which is slightly larger than $T_c$ obtained in $N_f=2+1$ lattice simulations. Also, in order to allow for a crossover phase transition for $\mu=0$ (which in our description corresponds to a second order transition) with $g$, $\lambda \sim 1$ we need that $g^2 > \lambda$. Since the effective potential is written as an expansion in powers of $a/T$ we need that this ratio is smaller than 1. From Eqs.~(\ref{relation}) and~(\ref{massvac}) the coupling constants are proportional to $m_\sigma$ which, from the above conditions, restricts the analysis to considering not too large values of $m_\sigma$. Since the purpose of this work is not to pursue a precise determination of the couplings but instead to call attention to the fact that the proper treatment of screening effects allows the linear sigma model to provide solutions for the CEP, we consider small values for $m_\sigma$. Given that $\sigma$ is anyhow a broad resonance, in order to satisfy the above requirements let us take for definitiveness $m_\sigma = 300$ namely, close to the two-pion threshold. Therefore, the allowed values for the couplings $\lambda$ and $g$ are restricted by
\bea
   \sqrt{\frac{\lambda}{2}+\frac{N_fN_cg^2}{6}} = 0.77.
\label{allowed}
\eea
Equation~(\ref{allowed}) provides a relation between $\lambda$ and $g$. A possible solution consistent with the above requirements is given as an illustration by $\lambda=0.2$, $g=0.71$. The corresponding phase diagram thus obtained is shown in Fig.~\ref{fig1}.
%%%%%%%%%%%%%%%%%%%%%%%%%%%%%%%%%%%
%%%%%%%%%%%%%%%%%%%%%%%%%%%%%%%%%%%

Note that for small $\mu$ the
phase transition is second order. In this case the
(pseudo)critical temperature is determined from setting the second
derivative of the effective potential in Eq.~(\ref{Veff-mid}) to
zero at $v=0$. When $\mu$ increases, the phase transition becomes
first order. The critical temperature is now computed by looking
for the temperature where a secondary minimum for $v\neq 0$ is
degenerate with a minimum at $v=0$. In both of these cases, from the detailed analysis, we
locate the position of the CEP as
$(\mu^{\mbox{\tiny{CEP}}}/T_c,T^{\mbox{\tiny{CEP}}}/T_c)\sim(1.2,0.8)$,
which is in the same range as the CEP found from lattice inspired
analyses~\cite{mathlattice}. Note also that the phase transition curve is essentially flat
close to the $T$ axis.

\section{Conclusions}

In conclusion, we have shown that it is possible to obtain values for the couplings that allow to locate the CEP in the region found by mathematical extensions
of lattice analyses. Since the linear sigma model does not 
have confinement we attribute this location to the adequate description
of the plasma screening properties for the chiral symmetry
breaking at finite temperature and density. 
%These properties are
%included into the calculation of the effective potential through
%the boson's self-energy and in the determination of the allowed
%range for the coupling constants through the observation that the
%thermal boson masses vanish at the phase transition for $\mu=0$. These observations
%determine a relation between the model parameters which is put in
%quantitative terms by taking physical values for $T_c$ from
%lattice calculations and for $a$ from the vacuum boson masses. 
Magnetic field effects can be included
in the description~\cite{Ayala4} both into the effective potential and into the behavior of the couplings, at lowest order. These last corrections lead to a decreasing of the couplings with the field strength. This decrease can be understood in general terms since the magnetic field produces a dimensional reduction whereby the virtual particles that make up the vacuum are effectively constrained to occupy Landau levels and thus restrict its motion to planes. This produces that charged virtual particles lie closer to each other and thus, because of asymptotic freedom, reduce the strength of the interaction. This happens no matter how weak the external field may be. We have found also that as the field strength increases, the CEP's location moves to lower values of $\mu$ and of $T$ and that in fact there is a value for this field where the CEP reaches the $T$-axis where the first order phase transitions remain for larger values of the field. These findings will be reported elsewhere shortly. We believe this description will
play an important role in determining the location of the CEP also in
QCD with and without magnetic fields.

\section*{Acknowledgments}

Support for this work has been received in part from UNAM-DGAPA-PAPIIT grant number 101515 and from CONACyT-M\'exico grant number 128534.

\end{document}